\documentclass[%
 aip,
 jmp,%
 amsmath,amssymb,
preprint,%
]{revtex4-1}
\usepackage{subfigure}
\usepackage{bm}
\usepackage{amssymb}
\usepackage{amsmath}
\usepackage{pxfonts}
\usepackage{latexsym}
\usepackage{setspace}
\usepackage{accents}
\usepackage{graphicx}
\usepackage{dcolumn}
\usepackage{bm}
\providecommand\bnabla{\boldsymbol{\nabla}}

\begin{document}

\preprint{AIP/123-QED}

\title[Fluid-Structure Interaction in Deformable Microchannels]{Fluid-Structure Interaction in Deformable Microchannels}

\author{Debadi Chakraborty}
 \altaffiliation[Also at ]{Department of Mathematics and Statistics, The University of Melbourne, Victoria~3010, Australia.}
\author{J. Ravi Prakash}%
\affiliation{Department of Chemical Engineering, Monash University, Melbourne, VIC~3800, Australia}

\author{Leslie Yeo}
\altaffiliation[Also at ]{Micro/Nanophysics Research Laboratory, School of Electrical and Computer Engineering, RMIT University, Melbourne, VIC~3000, Australia.}
\author{James Friend}
\altaffiliation[Also at ]{Micro/Nanophysics Research Laboratory, School of Electrical and Computer Engineering, RMIT University, Melbourne, VIC~3000, Australia.}
\affiliation{Department of Mechanical Engineering, Monash University, Melbourne, VIC~3800, Australia
}%

\date{\today}

\begin{abstract}
A microfluidic device is constructed from PDMS with a single channel having a short section that is a thin flexible membrane, in order to investigate the complex fluid-structure interaction that arises between a flowing fluid and a deformable wall. Experimental measurements of membrane deformation and pressure drop are compared with predictions of two-dimensional and three-dimensional computational models which numerically solve the equations governing the elasticity of the membrane coupled with the equations of motion for the fluid. It is shown that the two-dimensional model, which assumes a finite thickness elastic beam that is infinitely wide, approximates reasonably well the three-dimensional model, and is in excellent agreement with experimental observations of the profile of the membrane, when the width of the membrane is beyond a critical thickness, determined to be roughly twice the length of the membrane.
\end{abstract}

\maketitle

\section{\label{sec:Introduction}Introduction}
The fabrication of microfluidic devices from soft polymers or elastomers has gained considerable interest in the last decade. The attractiveness of using these soft materials, in particular, stems from the ability to tailor
the polymer's physicochemical properties specifically for a given application, the lower material and
fabrication costs which allows the possibility for disposable devices, and the durability of polymer-based
materials compared to the brittleness of conventional hard materials such as silicon and glass~\citep{ng02}.
Moreover, soft polymers such as polydimethysiloxane (PDMS)~\citep{james10} offer excellent optical transparency,
gas permeability and biocompatibility, vital for on-chip cell culture and comprising a large proportion
of microfluidic devices for cellomics, drug screening and tissue engineering~\citep{yeo11}.

The bonding strength, ability to mold at even nano-scale, biocompatibility, transparency and flexibility of these elastomeric substrates also make them ideal for fabricating microfluidic actuation structures. For
example, thin PDMS membranes have been employed as diaphragms or membrane interfaces for pneumatic actuation
and control in microchannels~\citep{vestad04,wang06,irimia2006b,huang2009} and substrates for biological
characterisation and manipulation in microdevices~\citep{thangawng2007, fuard08,solomon09} More sophisticated
microfluidic actuation structures have also been proposed, including multilayer and branched channel networks controlled
by elastomeric micropumps and microvalves~\citep{unger2000} for a variety of uses, for example, to spatiotemporally
control chemical gradients for chemotaxis studies on a microfluidic chip~\citep{irimia2006a}.

Fundamental studies in investigating the complex fluid-structural interaction arising from these flexible materials and the flow of the fluid within them, however, have not kept at the same rapid pace as the developments that have arisen. In fact, there have been no studies undertaken to investigate the flow through flexible channels at scales commensurate with microfluidic devices. To the best of our knowledge, even simple experiments~\citep{conrad69,brower75, bertram82, bertram86,
bertram87, bertram90, bertram91,bertram97, bertram99, bertram03} and theoretical studies ~\citep{luo95,luo96,hazel03,
HeilJensen2003b,Marzo2005,Liu2009} undertaken to investigate Newtonian flows through macroscopic deformable tubes have
yet to be reproduced at the microscale, where channel dimensions are of the order of 10--100 $\mu$m and the Reynolds number
is typically of the order of unity or below, typically two or more orders of magnitude smaller than the \cal{O}(10 mm) channel dimensions
and \cal{O}(100) or greater Reynolds numbers examined in these studies. The length scale at which fluid flow occurs in microfluidic devices
is entirely different from the large-scale flows. Fluid flowing in a conventional
microfluidic channel with characteristic length scale in the
sub-millimeter range, is identified by low velocity and hence small
Reynolds numbers. It is widely acknowledged that the experimental observations conducted in \emph{macroscale} channels
can be well predicted by the Navier-Stokes equation. However, to the best of our knowledge, there is no evidence in the literature of the use of a fluid-structural interaction theory to a collapsible microchannel. Fluid flow within a flexible structure is regulated by the stresses
imposed upon the structure by both the fluid and any external
forces. Thus, the rheological properties of both the fluid and the
structure significantly influence the fluid flow in the system. The
viscous stresses and fluid pressure exerted on the
boundaries of the flexible wall cause its deformation. Due to the deformation of the flexible structures, the flow domain and flow field alters and gives rise to an intricate fluid-structure interaction problem which requires the solution of a
free-boundary problem. Thus in order to understand fluid-structural interaction phenomenon at the microscale, successful development of fabricated microfluidic
devices as well as the implementation of a fluid-structural interaction theory associated with the microscale is necessary.

Given this motivation, we therefore carry out a fundamental investigation of the fluid flow in a deformable {\em micro}channel.
Specifically, we compare deformation profiles from experiments carried out on a custom fabricated microfludics with a flexible membrane section with that predicted by a two-dimensional finite element model that solves
the coupling between the membrane deformation and the fluid flow. We fabricated the structure by casting PDMS into a 200 $\mu$m high and 29 mm long microchannel. Membrane deformation was controlled in the experiment by the level of air pressure introduced
via a pressure chamber from the side of the membrane opposing the channel.

The rest of the article is thus organised as follows. We first formulate the numerical model and discuss the solution methodology in Sec. \ref{sec:FEMmicro}. The
fabrication and design of the deformable microchannel and the experimental methodology is subsequently described in Sec. \ref{sec:methodmicro}.
A comparison between the results obtained from both the experiments and numerical simulation then follows in Sec. \ref{sec:RDmicro},
after which we summarise our conclusions in Sec. \ref{sec:conclusionmicro}.

\section{\label{sec:FEMmicro} Numerical Simulation and Solution Methodology}

\subsection{\label{sec:2DFEMFSI} Two-Dimensional Finite Element Model for Fluid-structure Interaction (2D-FEM-FSI)}
\begin{figure}
\begin{center}
       \includegraphics[width=10.0cm]{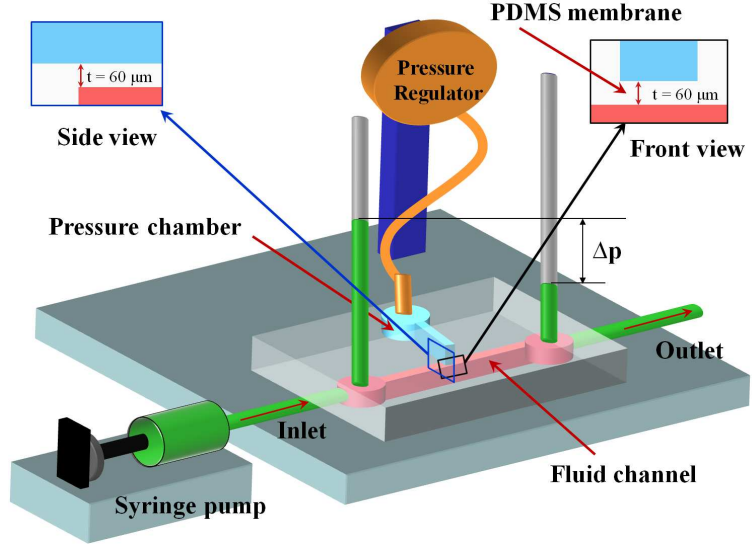}
       \begin{spacing}{1.5}
        \caption{\small  \label{mainfig1} Schematic illustration of the experimental setup.}
         \end{spacing}
\end{center}
\end{figure}

\begin{figure}
\begin{center}
       \includegraphics[width=14.0cm]{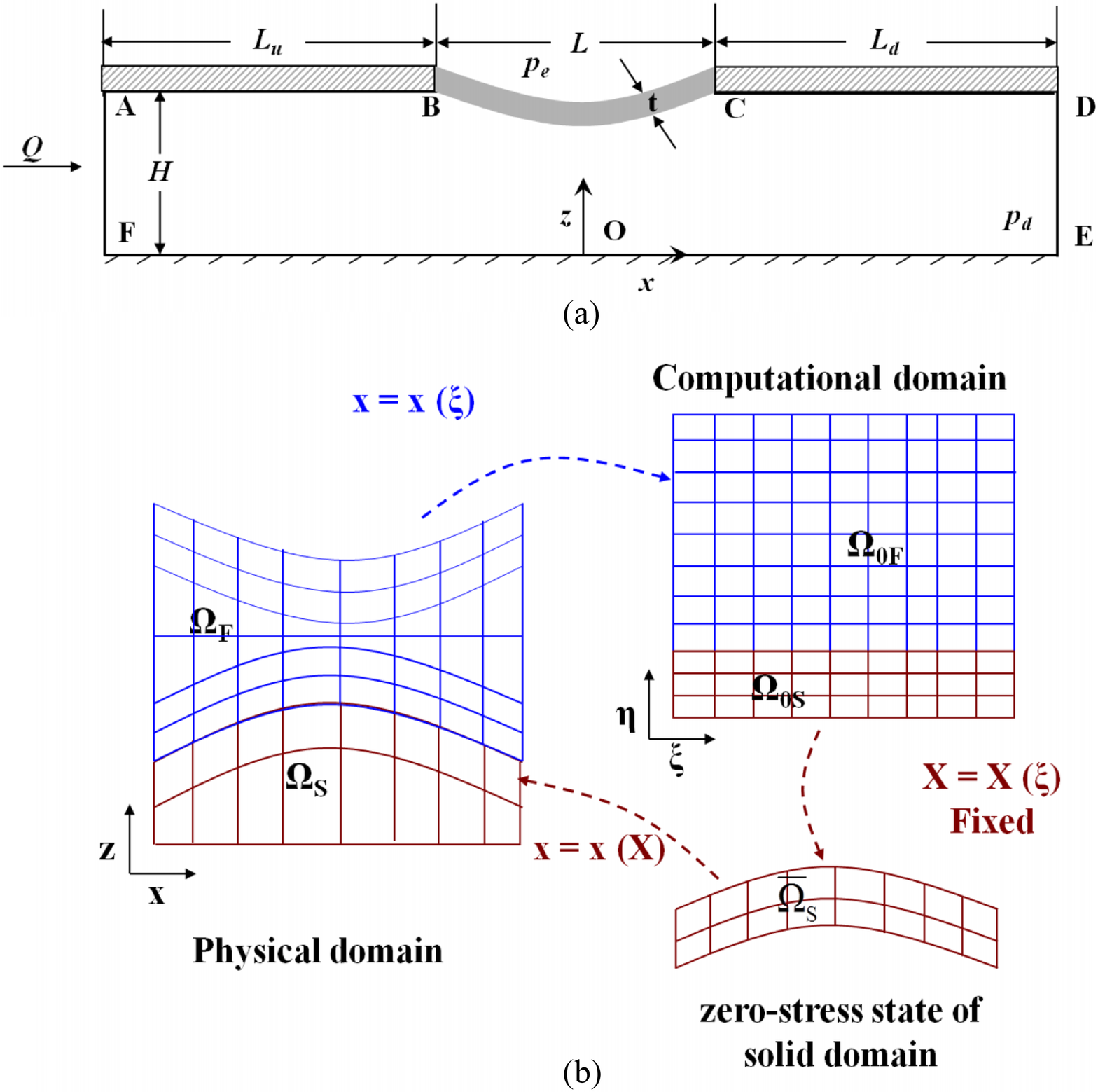}
       \begin{spacing}{1.0}
        \caption{\label{mainfig2}(a) Geometry of the two-dimensional deformable microchannel defining the solution space for the numerical simulation. (b) The solution strategy
        for the coupling between the fluid and solid domains that involves a free boundary problem is handled by mapping the physical fluid domain to
        a reference computational domain ($\Omega_F \rightarrow \Omega_{0F}$) and the physical solid domain to a reference zero-stress domain
        ($\Omega_S \rightarrow \overline{\Omega}_S$). We note that the solution of the elasticity equations itself constitutes a mapping from
        the zero-stress configuration ($\Omega_S \rightarrow \overline{\Omega}_S$), and is consequently not solved separately. The mapping from computational domain ($\Omega_{0S}$) to the zero-stress
configuration ($\overline{\Omega}_S$) is known and it only requires
a change of domain of integration. }

         \end{spacing}
\end{center}
\end{figure}

To match the geometry of the microchannel and the flexible membrane that spans the channel {\em width} in the experimental design (discussed subsequently in Sec. \ref{sec:methodmicro}) shown in
Fig. \ref{mainfig1}, we consider a two-dimensional
model of the experimental setup as shown in Fig.~\ref{mainfig2}(a), in which fluid flows through a section of the microchannel with height $H$, along that, a short segment of elastic membrane BC with thickness $t$ and length $L$ spanning the {\em width} of the channel exists on one side. Whilst the
sidewalls of the channel prior to and after the membrane section, AB and CD with lengths $L_u$ and $L_d$, respectively, are considered rigid,
the membrane is allowed to deform under an external pressure $p_e$ as measured in the air pressure chamber. To mimic the
experimental geometry, we set $H = 200\ \mu$m, $L_u = 14$~mm, $L = 1$~mm, $L_d = 14$~mm and $t = 60\ \mu$m. Here, the $x$-axis spans the channel length whereas the $z$-axis denotes the height of the channel with origin at point O.

In the absence of body forces, assumed negligible at the microscopic scales considered, the equations of motion for steady, incompressible flow
are specified by the continuity and momentum conservation, respectively:
\begin{equation}
\boldsymbol{\nabla}\boldsymbol{\cdot}\boldsymbol{v}=0,
\label{continuity}
\end{equation}
\begin{equation}
\rho\,\boldsymbol{v}\boldsymbol{\cdot}\bnabla\boldsymbol{v}=\bnabla\boldsymbol{\cdot}(-p\, \textbf{I}+\bm{\tau}),
\label{momentum}
\end{equation}
where $\rho$ is the density of the fluid, $\boldsymbol{v}$ is the liquid velocity field, $p$ the liquid pressure and $\bm{\tau}$ the viscous
stress tensor; $\textbf{I}$ represents the identity tensor. For a Newtonian fluid, $\bm{\tau} = 2\eta\textbf{D}$, where
$\eta$ is the viscosity of the liquid and $\textbf{D} =\frac{1}{2}(\bnabla\boldsymbol{v}+\bnabla\boldsymbol{v}^{\rm T})$ is the
strain rate tensor.

Given the deformability of the membrane under external pressure, the system comprises a free boundary problem in which the fluid
flow and the solid domain that constitutes the membrane are coupled. As we are interested in the steady flow, the inertia of the solid component is not affecting the overall dynamics of the system. We therefore employ a solution
strategy that continuously maps the fluid and solid domains $\boldsymbol{x}=\boldsymbol{x}(\bm{\xi})$, both unknown {\em a priori}, to arbitrary reference domains
following the approach of \citet{carvalho97}, as illustrated schematically in Fig.~\ref{mainfig2}(b).
Here, the physical and reference computational domains are parameterised by the position vector $\boldsymbol{x}$ and $\bm{\xi}$, respectively, and
$\boldsymbol{X}$ represents the position in the reference stress-free domain. The physical fluid domain $\Omega_F$
is mapped by elliptic mesh generation to a reference computational domain $\Omega_{0F}$, where Eqs.~(\ref{continuity}) and (\ref{momentum})
are solved. Due to the complexity in the geometry, the physical domain cannot be mapped to a simpler, quadrangular reference domain. Instead,
it is more convenient to subdivide the physical domain into subdomains and then map each subdomain into a separate subdomain of the reference
computational domain. Here we use a boundary-fitted finite element based elliptic mesh generation method~\citep{desantos91,chris92,
benjamin94, pasquali02} which involves solving the following elliptic differential equation for the mapping:
\begin{equation}
\bnabla\boldsymbol{\cdot} \tilde{\textbf{D}}\boldsymbol{\cdot}\bnabla{\bm{\xi}} = 0,
\label{ellipt}
\end{equation}
where the dyadic $\tilde{\textbf{D}}$ is a function of $\bm{\xi}$ in a manner analogous to a diffusion coefficient, which controls the
spacing of the coordinate lines \citep{benjamin94}.

The physically deformed solid domain constituting the flexible membrane $\Omega_S$, on the other hand, is mapped to a reference domain, that for convenience, we choose as a hypothetical zero-stress state where the stress tensor vanishes over the entire membrane (which may not and need not be physically realised). It is in this stress-free domain $\overline{\Omega}_S$ where the elasticity equations governing the deformation of the solid membrane are solved, although the solution of these equations itself constitutes a
mapping from the zero-stress configuration $\overline{\Omega}_S$ to the deformed domain $\Omega_{S}$. The
mapping from the computational domain ($\Omega_{0S}$) to the zero-stress
configuration ($\overline{\Omega}_S$) is known and only requires a change of the domain of integration.

In the reference stress-free domain $\overline{\Omega}_S$, the equilibrium equation that governs the solid if acceleration and body forces can be
neglected, reads
\begin{equation}
\bnabla_{\boldsymbol{X}}\boldsymbol{\cdot}\textbf{S}=0,
\label{cauchy-stress}
\end{equation}
where $\textbf{S}$ is the first Piola--Kirchhoff stress tensor. We note that this is related to the original deformed state of the solid, i.e., the
physical solid domain, through Piola's transformation to the Cauchy stress tensor $\bm{\sigma}$ by
\begin{equation}
\textbf{S}=\textbf{F}^{-1}\boldsymbol{\cdot}\bm{\sigma},
\end{equation}
where,
\begin{equation}
\textbf{F}=\frac{\partial \boldsymbol{x} }{\partial \boldsymbol{X} }
\end{equation}
is the deformation gradient tensor, which relates the undeformed state $\boldsymbol{X}$ = $(X, Y, Z)$ to the deformed state
$\boldsymbol{x}$ = $(x, y, z)$. Closure to the above is obtained through a constitutive relationship that relates the Cauchy stress tensor with
the strain. For a neo-Hookean material, this takes the form
\begin{equation}
\bm{\sigma}=-\pi^* \, \textbf{I}+G\textbf{B},
\label{NH}
\end{equation}
where $\pi^*$ is a pressure-like scalar function, $G$ is the shear modulus and $\textbf{B} = \textbf{F}\cdot\textbf{F}^{T}$.
$\textbf{B}$ is the left Cauchy-Green deformation tensor.

The equations governing the fluid motion and the solid deformation above are subject to the following boundary conditions:
\begin{enumerate}
\item No slip boundary conditions apply on the rigid walls, i.e., $\boldsymbol{v} = \boldsymbol{0}$ on $-(L_u+ L/2)\le x \le(L_d+ L/2)$ when $z = 0$ and $-(L_u+ L/2)\le x \le -L/2$ and $L/2 \le x \le (L_d+ L/2)$ when $z = H$.
\item Zero displacements are prescribed at the left side and right side of the solid, i.e., $\boldsymbol{x}$ = $\boldsymbol{X}$ on $H \le z \le H + t$ when $x = -L/2, L/2$.
\item At the upstream boundary ($0 \le z \le H$, $x = -(L_u+ L/2)$), a fully-developed velocity profile is specified, i.e., $v_z = 0$ and $v_x =U_0 {f(z/H)}$ where $U_0$ is the
average inlet velocity.
\item At the downstream boundary ($0 \le z \le H$, $x = (L_d+ L/2)$), the fully-developed flow boundary condition is imposed, i.e.,
$\boldsymbol{n}\boldsymbol{\cdot}\bnabla\boldsymbol{v = 0}$.
\item A force balance and a no-penetration condition are prescribed at the interface between the liquid and solid domain:
\begin{equation}
\boldsymbol{n} \boldsymbol{\cdot} \bm{\tau} = \boldsymbol{n} \boldsymbol{\cdot} \bm{\sigma}\quad \text{and }\quad
 \boldsymbol{u} = \boldsymbol{v} = \boldsymbol{0},
\label{interface}
\end{equation}
where $u$ is the velocity of the solid and $\boldsymbol{n}$ is the outward unit vector normal to the deformed solid surface.
\item A force balance is prescribed at the top surface.
\begin{equation}
\boldsymbol{n}.\bm{\sigma} = -p_{e}\boldsymbol{n}, \label{membrane}
\end{equation}
where $p_{e}$ is the external pressure.
\item We have compared our theory with the pressure drop only. So, the gauge pressure of the fluid at the downstream boundary is assumed negligible, i.e.., $p_d = 0$.
\end{enumerate}

The weighted residual form of Eqs. (\ref{continuity})--(\ref{cauchy-stress}), obtained by multiplying the
governing equations with appropriate weighting functions and subsequently integrating over the current domain, yields a large set of coupled
non-linear algebraic equations, which is solved subject to the boundary conditions specified above using Newton's method with
analytical Jacobian, frontal solver and first order arc length continuation in parameters~\citep{pasquali02,zevallos05,bajaj07,debadi10}.
The formulation of the fluid-structure interaction problem posed here follows the procedure introduced previously by~\citet{carvalho97} in their examination of roll cover deformation in coating flows. It turns out, however, that the weighted residual form of Eq.~(\ref{cauchy-stress})
used in their finite element formulation is incorrect. While insignificant for small deformations, this error leads to significant discrepancies when
the deformation is large. The weighted-residual equation is corrected here and validated in Appendix~\ref{ap:ftsm} against predictions using
a commercial finite element software ANSYS $11.0$~\citep{ansys} for the deformation of a simple beam fixed at its edges that we describe next.

\subsection{\label{sec:ANSYSmicro} Finite Element Model in ANSYS (2D/3D-ANSYS)}

We expect that the two-dimensional numerical simulation proposed above reasonably approximates a three-dimensional system when the
microchannel and hence membrane width is large compared to the height and length of the microchannel such that boundary effects at the
membrane edges can be neglected. To determine the limits of the microchannel width at which the two-dimensional approximation breaks down,
we carry out a three-dimensional finite element simulation that involves a plane-strain model for a compressible neo-Hookean solid since the incompressible neo-Hookean model described in Sec. \ref{sec:2DFEMFSI} is only valid for a two-dimensional geometry. For a given strain-energy density function or a elastic potential function of a neo-Hookean material, $\widehat{W}$,
\begin{equation}
\widehat{W}=\frac{G}{2}\left(I_1 - 3\right) + \frac{1}{d}\left(J - 1\right)^2,
\end{equation}
where $I_1 = \text{tr}(\textbf{C})$ is the first invariant of the right Cauchy-Green deformation tensor, $G$ the initial shear modulus of the
material, $d$ the material incompressibility parameter and $J=\text{det}(\textbf{F})$ the ratio of the deformed elastic volume
over the undeformed volume of material. The corresponding stress component is
\begin{equation}
\widehat{\textbf{S}}=2\frac{\partial \widehat{W} }{\partial \textbf{C} },
\label{eq:2PK}
\end{equation}
where $\widehat{\textbf{S}}$ is the second Piola-Kirchoff stress tensor and $\textbf{C}=\textbf{F}^{T}\cdot\textbf{F}$ is the right
Cauchy-Green strain tensor. If acceleration and body forces are negligible, the equilibrium equation for the deformed configuration is then
\begin{equation}
\bnabla_{\boldsymbol{x}}\boldsymbol{\cdot}\bm{\sigma}=0,
\label{cauchy-stress-micro}
\end{equation}
in which the Cauchy stress tensor is related to the second Piola-Kirchoff stress tensor in Eq.~(\ref{eq:2PK}) by
$\bm{\sigma}=J^{-1}\textbf{F}\cdot\widehat{\textbf{S}}\cdot\textbf{F}^{T}$.

A finite element simulation using ANSYS $11.0$~\citep{ansys} was employed to solve Eq.~(\ref{cauchy-stress-micro}) together with the following boundary conditions:
\begin{enumerate}
\item Zero displacements are prescribed at all the side edges of the solid.
\item A force balance is prescribed at the top surface.
\begin{equation}
\boldsymbol{n}.\bm{\sigma} = -p_{e}\boldsymbol{n}, \label{membrane2}
\end{equation}
where $p_{e}$ is the external pressure.
\item The pressure at the bottom surface is assumed zero.
\end{enumerate}
The simulations were carried out for the membrane geometry shown in Fig.~\ref{mainfig3}(a) in the absence of a fluid in order to compare the membrane deformation under an external pressure loading between a
two-dimensional and three-dimensional model. We verified with initial simulations of the full geometry mimicking the experimental setup, which
included the microchannel and pressure chamber, that the deformation was insensitive to the physical presence of the pressure chamber and a
microchannel, at least for the case when the fluid is absent and hence, it was sufficient to simulate a rectangular membrane of thickness $60\ \mu$m
and length 1 mm with fixed edges. The effect of varying the membrane width ($0.2, 0.5, 1.0, 2.0, 3.0$ and $4.0$ mm) was examined and
compared to a two-dimensional finite element model (infinite width assumption (Fig.~\ref{mainfig3}(a))) to determine the limits at which the
two-dimensional model breaks down. Here, the thin PDMS membrane is modelled as a nearly incompressible non-linear
neo-Hookean elastic beam with Poisson ratio $\nu=0.495$ and Young's modulus $E=2$ MPa, as determined from the nanoidentation tests in Appendix~\ref{c:nano}.
\section{\label{sec:methodmicro} Experimental Design and Methodology}

\begin{figure}
\begin{center}
       \includegraphics[width=16.0cm]{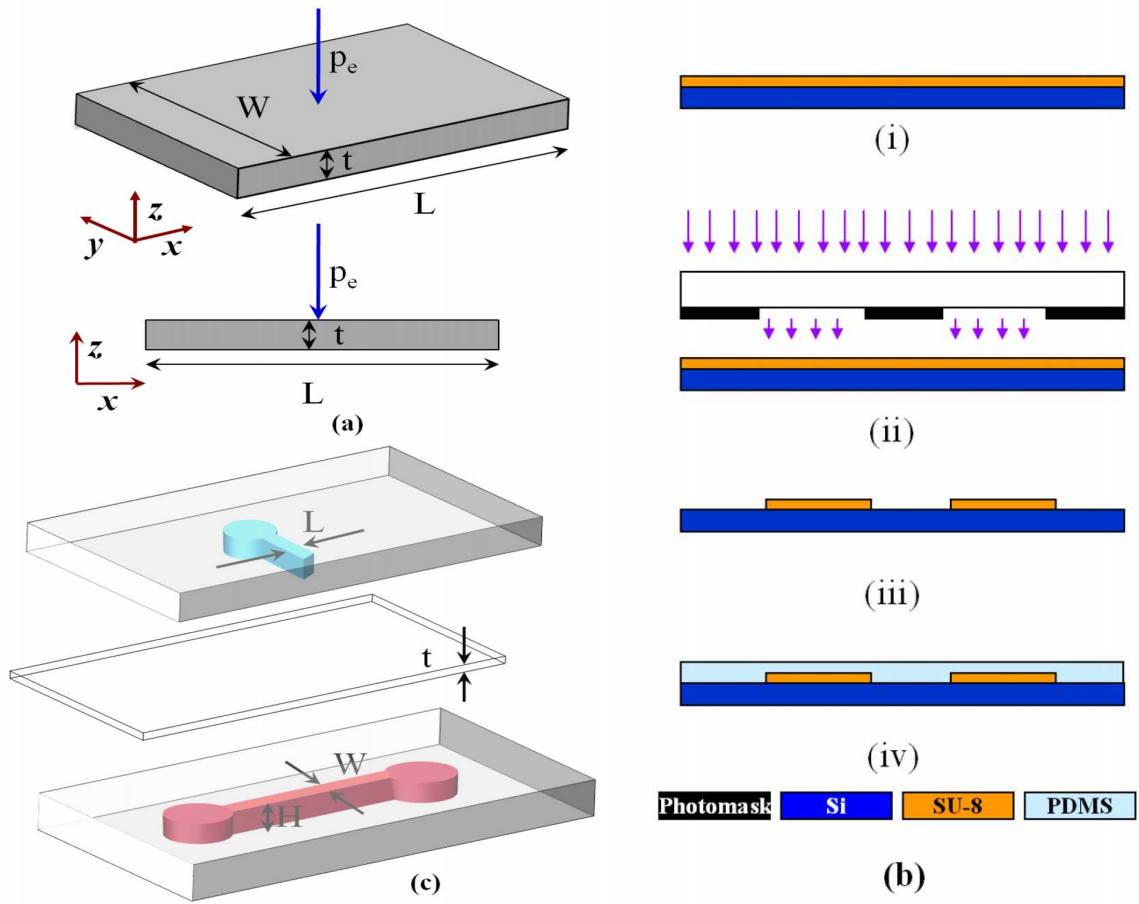}
       \begin{spacing}{1.5}
        \caption{\small (a) Schematic illustration of the membrane geometry indicating the definition of its width and length. (b) Schematic depiction of the main steps involved in the soft lithography
        procedure used to fabricate the PDMS deformable microchannel: (i) deposition of a SU-8 negative photoresist layer on a silicon
        wafer via spin coating, (ii) photoresist exposure to UV radiation through a photomask to polymerise the exposed regions, (iii) removal
        of uncrosslinked photoresist using developer solution to produce the replicated structure on the mould, and, (iv) inverse casting of the
        patterned mould in PDMS upon curing. (c) Exploded view schematic showing the microchannel (maroon) and pressure
        chamber (cyan) are cast separately in two PDMS layers and separated by an additional thin PDMS layer---the area of this layer which spans
        the microchannel depth $W$ constitutes the flexible membrane.}
        \label{mainfig3}
         \end{spacing}
\end{center}
\end{figure}
We fabricated the PDMS microfluidic channel using conventional soft lithography processes involving rapid prototyping and replica moulding typically
used elsewhere~\citep{duffy98,ng02,james10}, and schematically depicted in Fig.~\ref{mainfig3}(b). To prepare the replica moulds, we spin coat multiple layers of SU-8 (SU-8 2035, MicroChem,
Newton, MA, USA) negative photoresist onto clean silicon wafers, followed by pre-baking on a hot plate at 65$^{\text{o}}$C for 10 min and
subsequently 95$^{\text{o}}$C for 120 min to remove excess photoresist solvent. Repeated layering is required to prepare high aspect
ratio moulds with final thicknesses of up to 100--200 $\mu$m. The photoresist was then exposed to UV radiation at a wavelength of
350--400 nm for 60 s through a quartz photomask on which the device designs shown in Fig.~\ref{mainfig3}(c) are laser printed. This was followed
by a two-stage post-exposure bake at 65$^{\text{o}}$C for 1 min and 95$^{\text{o}}$C for 20 min to enhance the cross-linking in the exposed
portions of the SU-8. Finally, the wafer was developed to remove the photoresist in developer solution for 20 min and the mould dimensions
are verified by taking several measurements with a profilometer (Veeco Dektak 150, Plainview, NY; 1 $\AA$ maximum vertical resolution).

The base and curing agents of two-part PDMS (Sylgard 184, Dow Corning, Midland, MI, USA) is mixed in 10:1 ratio and kept in a vacuum
chamber to remove any bubbles generated during mixing.  The PDMS mixture is then poured over the mould and cured in an oven at 70$^0$C
for 2 hr. To ensure that the rigidity of the PDMS is maintained across the devices, the mixing ratio and curing procedure is strictly adhered to.
The PDMS channel replica was then peeled off the mould and inlet, outlet and pressure sensor ports were drilled into the structure. Finally, the
PDMS channel was oxidised in a plasma cleaner for 2 min and sealed by bonding against a flat PDMS substrate.

The microchannel and pressure chamber are cast in two separate PDMS layers and interleaved with an additional thin PDMS layer for the design shown in Fig.~\ref{mainfig3}(c). Microchannels with $0.2, 0.5, 1.0, 2.0, 3.0$ and $4.0$~mm widths were constructed with this technique. The thickness of the flexible membrane was fabricated to be $60\ \mu$m for all channel widths.

\begin{figure}
\begin{center}
\includegraphics[width=10.0cm]{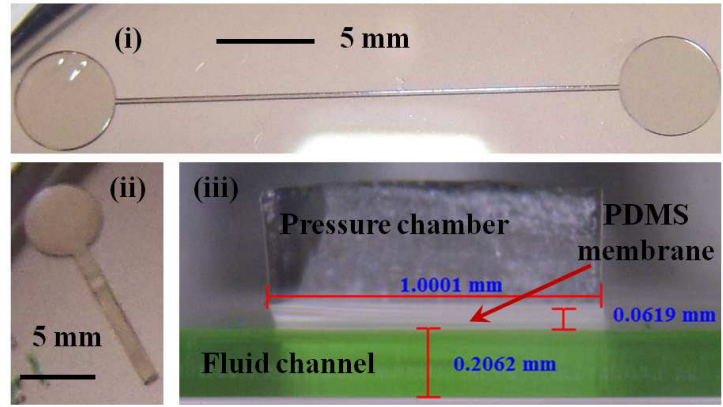}
\begin{spacing}{1.5} \caption{\small \label{mainfig4} Images of the patterned SU-8 mould and the PDMS
devices for the design depicted in Fig.~\ref{mainfig3}(c), panel (i) shows the SU-8 mould for the microchannel, panel (ii) shows that for the
pressure chamber, and panel (iii) shows a micrograph of the thin interleaving PDMS layer that constitutes the flexible membrane that separates the
microchannel and the pressure chamber. }
\end{spacing}
\end{center}
\end{figure}

\subsection{\label{sec:fsi} Deformation of the PDMS membrane with fluid flow}

The deformation of the thin PDMS membrane under an external air pressure (Precision pressure regulator-IR1020, SMC Pneumatics, Australia) applied to the chamber was measured visually using a microscope
and video camera (AD3713TB Dino-Lite Premier, AnMo Electronics, Hainchu, Taiwan; 640 x 480 pixel resolution and 200X maximum magnification). Nanoindentation experiments, on the other hand, were carried out to evaluate the values
of the Young's modulus for the PDMS membranes required in the numerical simulations, from which values in the range of $1.2$ to $2.2$ MPa were
obtained. A detailed description of the nanoindentation test results can be found in Appendix~\ref{c:nano}. Water and sucrose syrup were employed as the working fluid, which were driven through the microchannel at a constant flow rate using a syringe pump
connected to the channel inlet. To measure the pressure drop in the channel, we vertically mount capillary tube at the inlet and outlet as illustrated in Fig.~\ref{mainfig1} and determine the difference in the height across the fluid columns in the capillary tubes.

\section{\label{sec:RDmicro} Results and Discussion}

\subsection{\label{sec:2d3d} Comparison Between Two- and Three-Dimensional Models in the Absence of Flow}
\begin{figure}
\begin{center}
\includegraphics[width=16.0cm]{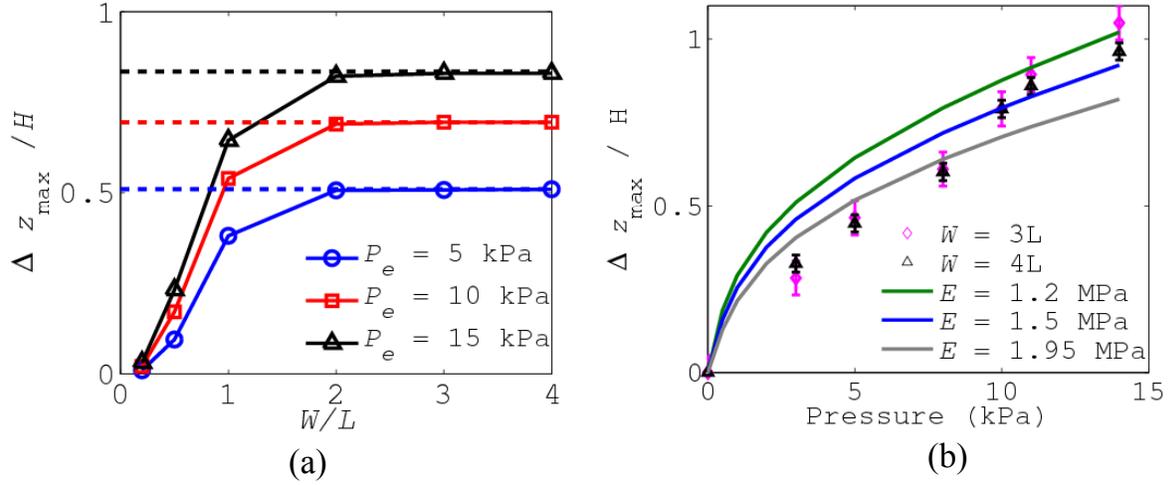}
\begin{spacing}{1.5} \caption{\small \label{mainfig5} (a) Maximum deformation $\Delta z_{\text{max}}$ of the lower surface
of the flexible membrane as a function
of its width $W$ for three different values of the applied external pressure $P_e$, as predicted by the 3D-ANSYS simulation described in  Sec. \ref{sec:ANSYSmicro}. The solid lines were added to aid visualisation. Also shown by the dashed lines is a
2D-ANSYS (infinite width assumption) finite element simulation, from which we observe the diminishing effect of boundary effects
associated with the pinned lateral edges in the three-dimensional model above a width of approximately 2 mm. (b) Beyond this critical width,
therefore, we verify that a two-dimensional simulation (2D-FEM-FSI) is a good approximation to capture the behaviour of that observed in experiments.}
\end{spacing}
\end{center}
\end{figure}

\begin{figure}
\begin{center}
\includegraphics[width=16.0cm]{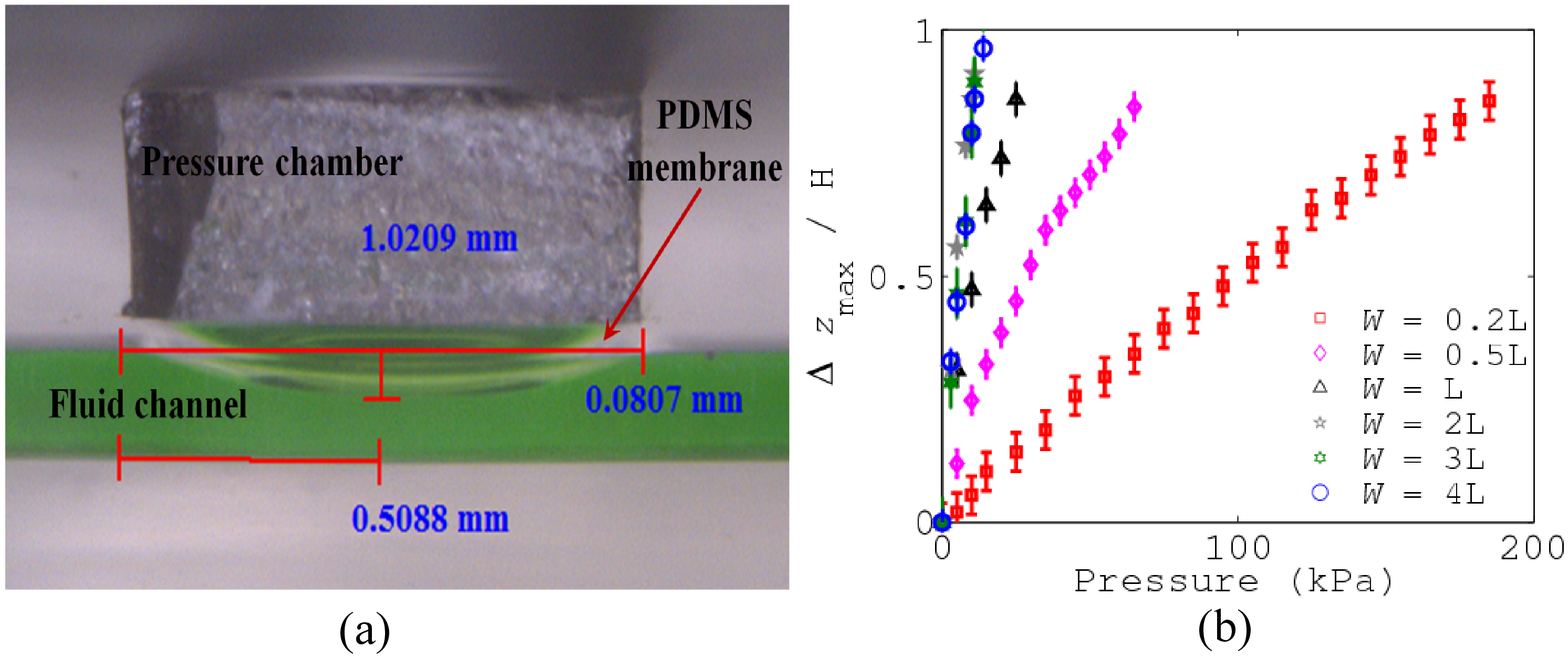}
\begin{spacing}{1.5} \caption{\small \label{mainfig6}(a) Micrographs
        showing the deformation of the thin flexible PDMS membrane under the application of an external pressure $p_e$  of $20$ kPa for the microchannel designs shown in Fig.~\ref{mainfig3}(b) with a channel width of approximately $0.5$ mm. (b) Maximum deformation of the membrane $\Delta z_{\text{max}}$, measured at the lowest point
        of the inflexion of its lower surface, as a function of the externally applied pressure for microchannels with varying widths.}
\end{spacing}
\end{center}
\end{figure}

Figure~\ref{mainfig5}(a) that depicts the maximum deformation
predicted by the two-dimensional and three-dimensional ANSYS simulations (2D/3D-ANSYS) described in Sec. \ref{sec:ANSYSmicro}, shows that the two-dimensional
model begins to deviate from the three-dimensional prediction below a membrane width of 2$L$, when boundary effects associated with edge
pinning on both sides can no longer be neglected. This is consistent with what we observe in the experiments where we measure the deformation of
the thin PDMS membrane under an external air pressure loading in the absence of fluid flow. Figure~\ref{mainfig6}(a) shows the deformed shape of the membrane under an externally applied pressure. The maximum deformation
of the membrane, measured at the lowest point of the inflexion of its lower surface, is extracted visually from similar micrographs and shown in
Fig.~\ref{mainfig6}(b) as a function of the applied external
 pressure for microchannels of different widths. The presented experimental data point is the statistical average of at least five values, with vertical bars indicating the range of the deviation. The experimental error resulted from the manual handling of the microscope and video camera, image analysis to extract the deformed shape of the membrane and manual measuring of the fluid column height in the capillary tubes. In agreement with the predictions of the numerical simulations, we see that
the deformation becomes independent of the microchannel (and membrane) width when it exceeds 2$L$, therefore suggesting that
boundary effects associated with the membrane pinning at the lateral edges in a three-dimensional model can be neglected and a two-dimensional
(infinite width) approximation suffices beyond this critical dimension.

The validity of the two-dimensional incompressible neo-Hookean model (i.e., 2D-FEM-FSI) is further verified against experimental data for microchannels with large widths above 2$L$.
Figure~\ref{mainfig5}(b) shows a comparison between the maximum deformation
measured in the experiments with that predicted by the two-dimensional model, in that, we observe agreement with the experimental data to be bounded
by the numerical predictions using two values of Young's modulus for the membrane. We note that the large deformation data is well predicted by a
lower value of the Young's modulus, whereas, better agreement with the small deformation data is captured using a larger value. Both lower and upper values
however fall within the $1.2$ to $2.2$ MPa range measured using the nanoindentation technique described in Appendix~\ref{c:nano}. In any case, the
result provides further evidence to suggest that the two-dimensional model is sufficient to capture the membrane deformation when it
is beyond a critical value of 2$L$, that is consistently predicted by both experiment and simulation.

\subsection{\label{sec:flow} Flow Experiments: Pressure Drop and Membrane Deformation}

\begin{figure}
\begin{center}
       \includegraphics[width=10.0cm]{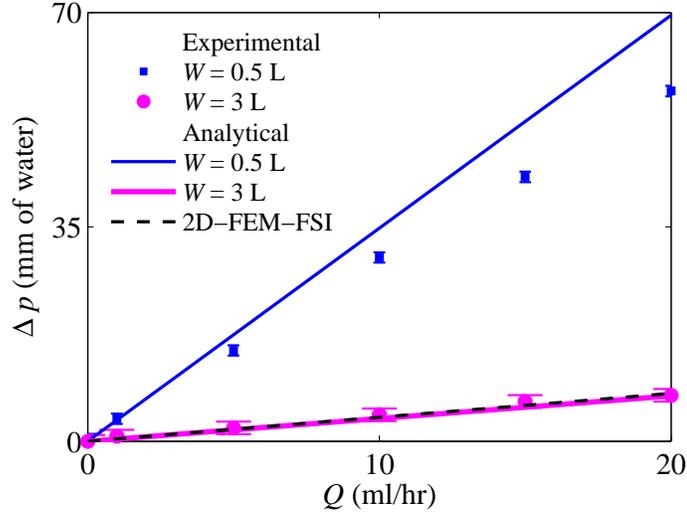}
       \begin{spacing}{1.5}
        \caption{\small  \label{mainfig7} Relationship between the pressure drop and flow rate across a deformable microchannel with two different
        channel widths for the wider channel. It can be seen that the experimental observations (symbols) match well with the predictions afforded by the finite element
        simulation (i.e., 2D-FEM-FSI) described in Sec. \ref{sec:2DFEMFSI} (dashed line) as well as
        that for a rigid rectangular microchannel (Eq.~(\ref{flowrate}); solid line). For the narrower channel, however, we observe
        that the analytical solution for the rigid microchannel overpredicts the pressure drop at higher flow rates.}
         \end{spacing}
\end{center}
\end{figure}

Figure~\ref{mainfig7} shows the pressure drop $\Delta p$ as a function of the flow rate $Q$ obtained from both experimental measurements and that predicted by the finite element model (2D-FEM-FSI) described in Sec. \ref{sec:2DFEMFSI}. Also shown is the pressure drop and flow rate
relationship for
two-dimensional fully-developed viscous flow through a long and {\rm rigid} rectangular microchannel, for which, it is possible to obtain an analytical solution
if the channel height $H$ and width $W$ are small compared to the channel length $L_c$---the solution for longitudinal velocity takes the
form~\citep{mortensen2005}
\begin{equation}
v_x=\frac{4 H^2 \Delta p}{\pi^3\eta L_c}\sum_{n, odd}^{\infty}\frac{1}{n^3}\left[1-\frac{\cosh(n\pi y / H)}{\cosh(n\pi W / 2H)}
\right]\sin \left(\frac{n\pi z}{H} \right).
\label{vprofile}
\end{equation}
Integrating along the width and height of the channel then gives the required pressure drop and flow rate relationship:
\begin{equation}
Q=\frac{H^3 W \Delta p}{12\eta L}\left[1 - \sum_{n, odd}^{\infty}\frac{192 H}{n^5 \pi^5 W} \tanh \left( \frac{n\pi W}{2H} \right)\right].
\label{flowrate}
\end{equation}
We observe very good agreement between the experimental pressure drop and flow rate relationship and that predicted by both the analytical solution
for a rigid microchannel and the 2D-FEM-FSI for the
case of the wide channel, that lends further support that the two-dimensional model constitutes a good approximation when the channel, and hence,
membrane is sufficiently wide, such that, three-dimensional effects, such as, the pinned boundaries at the sidewalls can be neglected. The good agreement with the analytical solution, that does not account for the flexible membrane also suggests that the effect of the deformation on the pressure drop is small, and hence can be neglected. This is however not true for the case for small channel widths where we observe a departure from the rigid channel prediction at larger flow rates. Compared to a rigid channel, the ability of the membrane to deform in a deformable channel gives rise to smaller effective cross-sectional areas that in turn, lead to faster velocities in order that continuity is satisfied. Consequently, a lower pressure drop is required to sustain the same
flow rate compared to a rigid channel with a larger cross-sectional area.
\begin{figure}
\begin{center}
\includegraphics[width=16.0cm]{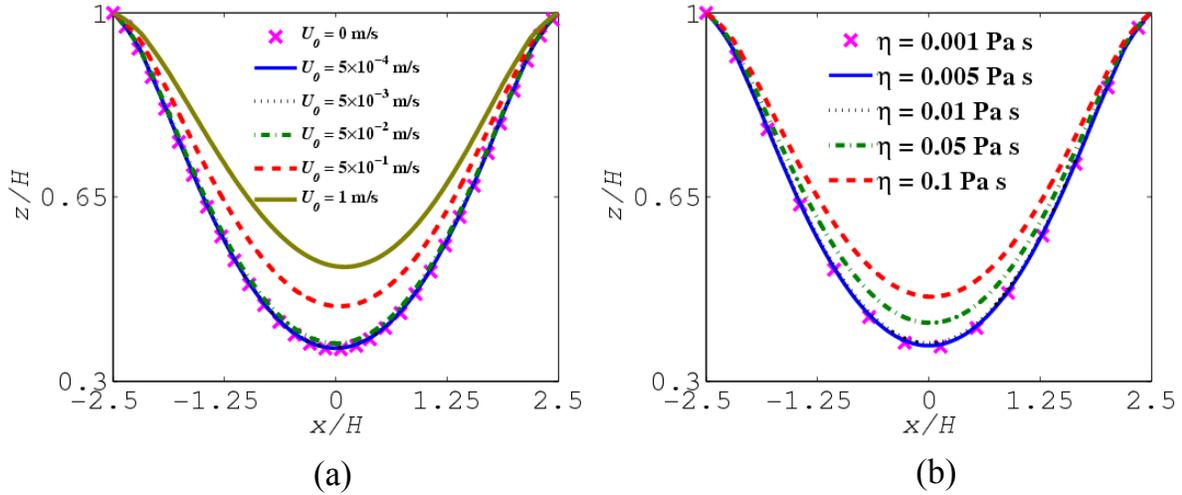}
       \begin{spacing}{1.5}
        \caption{\small  \label{mainfig8} Shape of the membrane profile as predicted by the finite element model (i.e., 2D-FEM-FSI) described in Sec. \ref{sec:2DFEMFSI} for varying (a) average inlet velocity and (b) fluid viscosity. The rest of the parameters are $p_e$ = $8$ kPa,
        $E$ = $1.95$ MPa, $t = 60\ \mu$m, $H = 200\ \mu$m and $W = 3$ mm. In
        (a), $\eta = 0.001\ {\rm Pa}\cdot{\rm s}$ and in (b), $U_0 = 5 \times 10^{-3}$.}
         \end{spacing}
         \end{center}
\end{figure}

Figure~\ref{mainfig8}(a) shows profiles of the deformed membrane shape
under a specified external pressure for varying flow rates (and hence corresponding average inlet velocities $U_0$), as predicted by the 2D-FEM-FSI
simulation described in Sec. \ref{sec:2DFEMFSI}. We observe no significant deformation
of the membrane below $U_0 = 5 \times 10^{-2}$ m/s corresponding to a flow rate of $Q = 110$ ml/hr. This is confirmed in our experiments where we do not see any
measurable changes in the membrane shape at these flow rates. Restrictions in the maximum flow rate due to experimental limitations, nevertheless, did
not allow us to access flow rate regimes where the numerical simulation predicts observable changes in the membrane shape.

\begin{figure}
\begin{center}
\includegraphics[width=16.0cm]{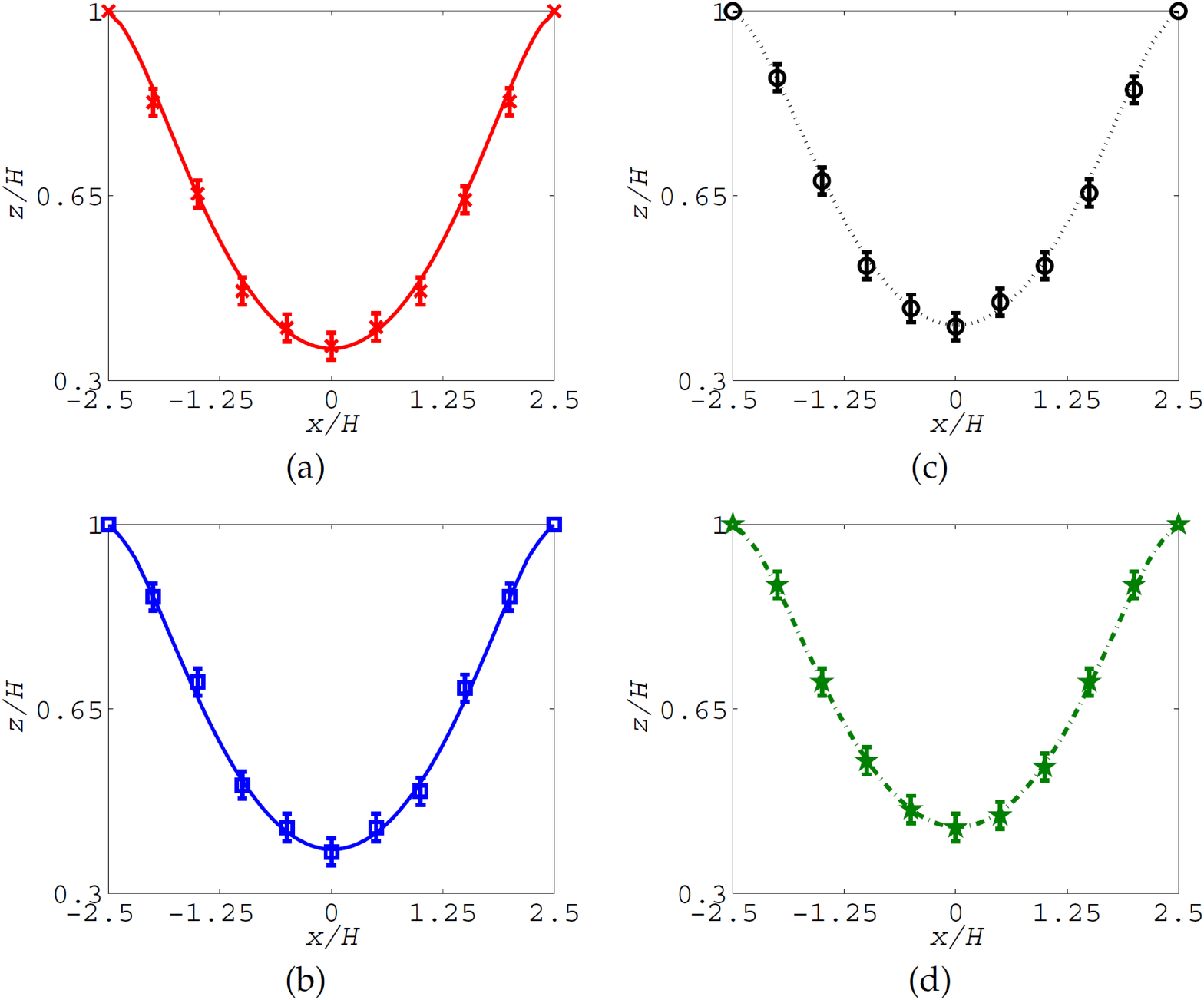}
\begin{spacing}{1.0}
\caption{\small \label{mainfig9} Comparison of the shape of the membrane predicted by the 2D-FEM-FSI (lines) described in Sec. \ref{sec:2DFEMFSI} with that measured experimentally (symbols) for different inlet fluid velocities, (a) $U_0$ = 0, (b) $U_0$ = $2.2\times 10^{-3}$, (c) $U_0$ = $4.4\times 10^{-3}$ m/s and (d) $U_0$ = $6.6\times 10^{-3}$ m/s. The rest of the parameters are $p_e$ = $8$ kPa, $E$ = $1.95$ MPa, $t = 60\ \mu$m, $H = 200\ \mu$m, $W = 3$ mm and $\eta = 0.05\ {\rm Pa}\cdot{\rm s}$.}
\end{spacing}
\end{center}
\end{figure}

Fortunately, however, the membrane deformation is sensitive to the fluid viscosity, as shown by the profiles predicted by the numerical simulation
in Fig.~\ref{mainfig8}(b). The experiments were therefore repeated under the same conditions
but with sucrose syrup with a viscosity of $0.05\ {\rm Pa}\cdot{\rm s}$ in order to obtain measurable deformations in the membrane shape.
Figure~\ref {mainfig9} shows the close agreement between the shape of the membrane profile that is experimentally measured with that predicted by
the numerical simulation (i.e., 2D-FEM-FSI) described in Sec. \ref{sec:2DFEMFSI}, therefore inspiring confidence in the predictive capability of the proposed model.

\section{\label{sec:conclusionmicro} Conclusions}

To investigate the complex fluid-structural interactions between a deformable channel wall and the fluid that flows within it, we fabricate a microfluidic
device that constitutes a single channel out of PDMS with a short section comprising a thin flexible membrane. Experiments in which we measure the
membrane deformation and the pressure drop across the channel are complemented by the development of predictive computational models, in which, we numerically solve the equations of motion in the fluid coupled with the equilibrium equations governing the elasticity of the membrane. In particular,
we show that two-dimensional models can only be used to describe a three-dimensional system, when the width of the channel, and hence, the membrane is sufficiently large above a critical dimension, such that, boundary effects arising from the pinning of the membrane to the channel walls at its lateral edges can be neglected---the 2$L$ threshold predicted by the simulations agrees well with our experimental observations. In addition, we find excellent agreement between the predictions of the deformed membrane shape under an externally applied air pressure using both two-dimensional and
three-dimensional models with that measured in experiments. We believe that the combination of these results, the predictive capability of the numerical
models developed, and the physical insight gleaned in this study would be useful in the design of polymer-based microfluidic devices, and, in particular, microactuation schemes such as the pneumatically-driven micropumps, micromixers, microvalves and microfilters  employing flexible polymer membranes that have grown increasingly popular over the last decade.
\begin{acknowledgments}
We thank Matheo Pasquali and Marcio Carvalho for providing us with their finite element code for simulating coating flows, which we have modified and adapted to this work. This work was supported by an award under the Merit Allocation Scheme on the NCI National Facility at the Australian National
University (ANU). The authors would also like to thank  VPAC (Australia), and SUNGRID (Monash University, Australia) for the
allocation of computing time on their supercomputing facilities. LYY is supported by an Australian Research Fellowship awarded
by the Australian Research Council under Discovery Project grant DP0985253. JRF is grateful for the MCN Tech Fellowship from the Melbourne Centre for Nanofabrication and the Vice-Chancellor's Senior Research Fellowship from RMIT University.
\end{acknowledgments}

\appendix

\section{Weighted Residual Form of the Equilibrium Equation $\bm{\nabla}_{\textbf{X}}\cdot\textbf{S}=\textbf{0}$}\label{ap:ftsm}

Here, we provide a correction to the weighted residual form of the equilibrium equation given by Eq.~(\ref{cauchy-stress}) derived by
~\citet{carvalho97}. The error in the original derivation, whilst insignificant for small deformations, leads to significant discrepancies when
the deformation is large.

The weak form of Eq.~(\ref{cauchy-stress}) is
\begin{equation}
\int_{\overline{\Omega}_S} \left(
\bm{\nabla}_{\textbf{X}}\cdot\textbf{S} \right) \phi \,
 \, d\overline{\Omega_S} = - \int_{\overline{\Omega}_S} \left( \bm{\nabla}_{\textbf{X}} \phi \cdot \textbf{S}
 \right) \, d\overline{\Omega}_S+\int_{\overline{\Gamma}_{S}} \phi \left(\textbf{N}\cdot\textbf{S}\right)
\, d\overline{\Gamma}_{S} = \textbf{0},
\label{Galerkin-solid}
\end{equation}
where $\overline{\Omega}_S$,  $\overline{\Gamma}_S$ and $\textbf{N}$ are the area, arc length and unit normal in the zero-stress configuration, respectively, and $\phi$ is a weighting function. When written in terms of Cartesian components, the
weighted residual form of this equation in the \emph{computational}
domain is
\begin{equation}
R^x_i=-\int_{\Omega_{S0}}\left[\frac {\partial \phi_i}{\partial
X}S_{Xx}+\frac {\partial \phi_i}{\partial Y} S_{Yx}\right]
|\textbf{J}^*| \,
d\Omega_{S0}+\int_{\Gamma_{S0}}\phi_i\left(\textbf{N}\cdot\textbf{S}\right)_x
\left(\frac{d\overline{\Gamma}_{S}}{d\Gamma_{S0}}\right)d\Gamma_{S0},
\label{x-position1}
\end{equation}
\begin{equation}
R^y_i=-\int_{\Omega_{S0}}\left[\frac {\partial \phi_i}{\partial
X}S_{Xy}+\frac {\partial \phi_i}{\partial Y} S_{Yy}\right]
|\textbf{J}^*| \,
d\Omega_{S0}+\int_{\Gamma_{S0}}\phi_i\left(\textbf{N}\cdot\textbf{S}\right)_y
\left(\frac{d\overline{\Gamma}_{S}}{d\Gamma_{S0}}\right)d\Gamma_{S0}.
\label{y-position1}
\end{equation}
Here, $\overline{\Omega}_{S0}$ and  $\overline{\Gamma}_{S0}$ is the
area and arc length in the computational domain, respectively,
$|\textbf{J}^*|$ is the Jacobian of the transformation from the
zero-stress configuration to the computational domain and
$\phi_{i}$ is a bi-quadratic weighting function. The components of the dimensional Piola-Kirchhoff stress tensor $\textbf{S}$
in terms of the \textit{dimensional} Cauchy stress tensor for a neo-Hookean material $\bm{\sigma} = - \pi^{*}\ \textbf{I} + G \, \textbf{B} $,
are
\begin{equation}
\begin{aligned}
S_{Xx}&=- \pi^{*} \, \frac{\partial y }{\partial Y}+ G \,
\frac{\partial x}{\partial X}, \quad S_{Yx} = \pi^{*} \,
\frac{\partial y }{\partial X}+  G \, \frac{\partial
x}{\partial Y}, \\
S_{Xy}&= \pi^{*} \, \frac{\partial x }{\partial Y}+ G \,
\frac{\partial y}{\partial X}, \quad S_{Yy} =- \pi^{*} \,
\frac{\partial x }{\partial X}+  G \, \frac{\partial y}{\partial Y}.
\label{MR2}
\end{aligned}
\end{equation}
where $\textbf{I}$ is the
identity tensor, $\pi^*$ is a pressure-like scalar function and
$\textbf{B}$ is the left Cauchy-Green tensor. In their finite-element formulation of the fluid-structure
interaction problem, \citet{carvalho97} (see also~\citet{carvalho96}) have used
\begin{equation}
R^x_i=-\int_{\Omega_{S0}}\left[S_{Xx}\frac {\partial
\phi_i}{\partial X}+S_{Xy}\frac {\partial \phi_i}{\partial Y}
\right] |\textbf{J}^*| \,
d\Omega_{S0}+\int_{\Gamma_{S0}}\phi_i\left(\textbf{N}\cdot\textbf{S}\right)_x
\left(\frac{d\overline{\Gamma}_{S}}{d\Gamma_{S0}}\right)d\Gamma_{S0}
\label{x-position2}
\end{equation}
and
\begin{equation}
R^y_i=-\int_{\Omega_{S0}}\left[S_{Yx}\frac {\partial
\phi_i}{\partial X}+S_{Yy}\frac {\partial \phi_i}{\partial Y}
\right] |\textbf{J}^*| \,
d\Omega_{S0}+\int_{\Gamma_{S0}}\phi_i\left(\textbf{N}\cdot\textbf{S}\right)_y
\left(\frac{d\overline{\Gamma}_{S}}{d\Gamma_{S0}}\right)d\Gamma_{S0}
\label{y-position2}
\end{equation}
in place of~Eqs.~(\ref{x-position1}) and~(\ref{y-position1}).
Essentially, the positions of the two components $S_{Yx}$ and $S_{Xy}$
have been interchanged.

\begin{figure}
\begin{center}
       \includegraphics[width=0.80\textwidth]{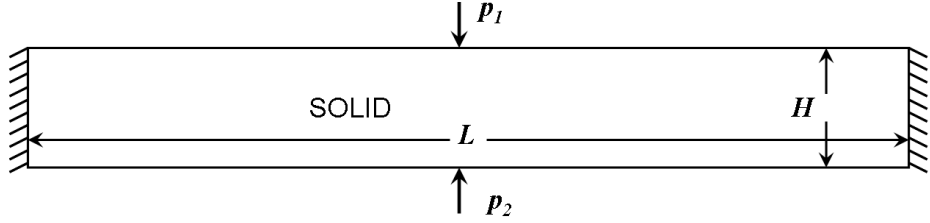}
\end{center}
        \caption{Geometry of the solid domain.}
        \label{appenAfig1}
\end{figure}

In order to establish the validity of Eqs.~(\ref{x-position1}) and~(\ref{y-position1}) and to demonstrate the
incorrectness of equations~(\ref{x-position2}) and~(\ref{y-position2}), we have examined the simple problem
of a beam fixed at the edges with uniform pressure applied on both the top and bottom of the beam,
as shown schematically in Fig.~\ref{appenAfig1}. Essentially, we compare the results of our computations using Eqs.~(\ref{x-position1}) and~(\ref{y-position1}) (labelled FEM-N), and Eqs.~(\ref{x-position2}) and~(\ref{y-position2}) (labelled FEM-C), with the results obtained with the finite element ANSYS
simulation for a plain-strain model described Sec. \ref{sec:ANSYSmicro}. In addition, we prescribe boundary conditions
in the form of zero displacements at the left and right edges of the beam, and a force balance at the top and bottom of the form,
\begin{equation}
\textbf{n} \cdot \bm{\sigma} = - p_{i} \, \textbf{n}, \quad (i = 1,2)
\label{membrane-JFM}
\end{equation}
where $\textbf{n}$ is the unit normal to the deformed solid surface, and $p_{1}$ and $p_{2}$ are the dimensional external
pressures on the top and bottom of the beam, respectively.

\begin{figure}
\begin{center}
       \includegraphics[width=9.0cm]{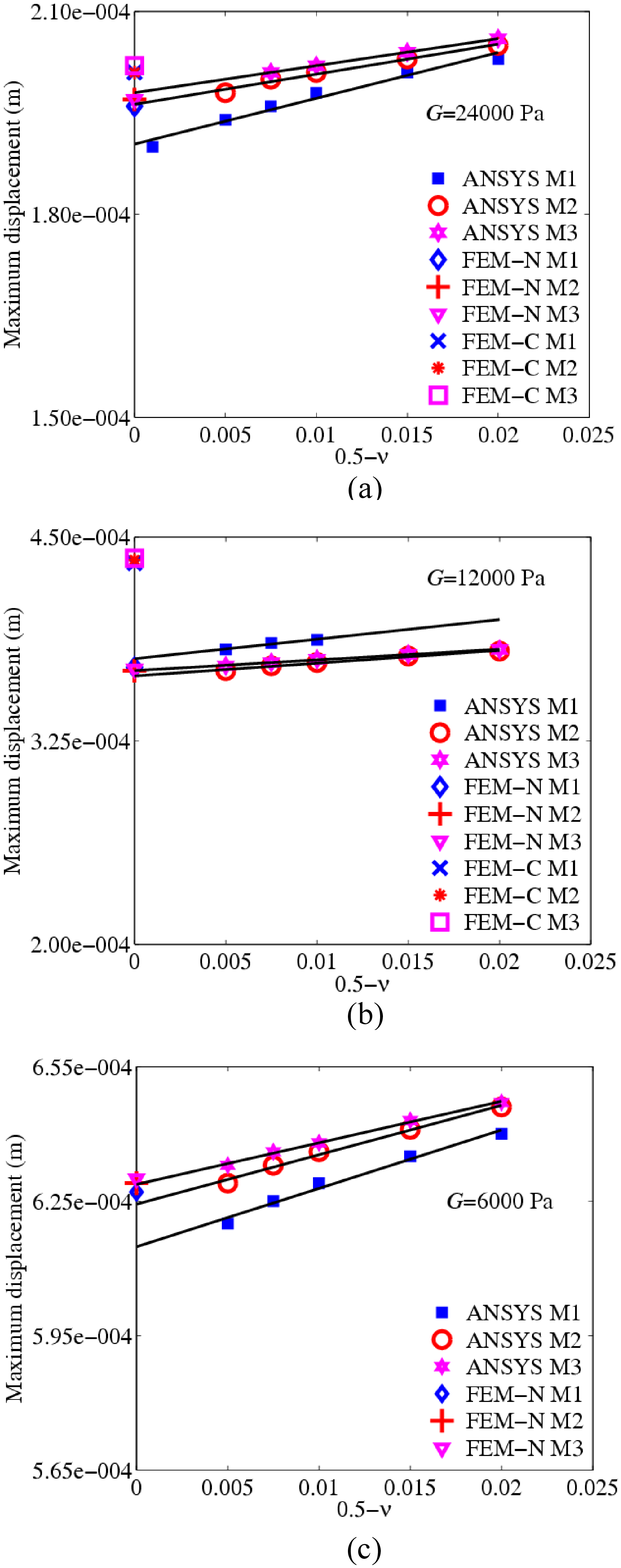}
\begin{spacing}{1.0}
 \caption{\small \label{appenAfig2} Comparison of formulations FEM-N and FEM-C with
 the ANSYS plain-strain model for three different values of
 $G$: (a) 24000 Pa,  (b) 12000 Pa, and  (c) 6000  Pa.}
\end{spacing}
\end{center}
\end{figure}
In units of height $H$, the length of the beam is set at $L = 5H$,
with $H$ = 10$^{-3}$ m. The external pressures have been chosen to
be $p_1 = 1.1$ N and $p_2 = 1.0$ N, and three different values
(6000, 12000 and 24000 Pa) have been used for the shear modulus $G$.
Computations have been performed with three different meshes (M1, M2
and M3) in order to examine mesh convergence. We note that the formulation of the fluid-structure interaction problem by
\citet{carvalho97} applies to the special case of an
\emph{incompressible} neo-Hookean material with a Poisson ratio $\nu
= 0.5$. On the other hand, the ANSYS plain-strain package is only
applicable to \emph{compressible} neo-Hookean materials.
Consequently, in order to carry out the comparison with the ANSYS simulations, we
have obtained predictions with several values of $\nu < 0.5$, and
extrapolated the results to $\nu = 0.5$.

Figure~\ref{appenAfig2} compares the maximum displacement of
the beam obtained with the FEM-N and FEM-C formulations with the
ANSYS plain strain model for the three different values of $G$.
In all three cases, mesh-converged results obtained with FEM-N are observed to agree with the extrapolated mesh converged
solution obtained with ANSYS. On the other hand, the mesh converged solution obtained
with FEM-C shows differences from the other two approaches. Indeed, while this difference is small at
$G = 24000$ Pa, and substantially larger at $G = 12000$ Pa, we are unable to obtain a converged solution with FEM-C
at $G = 6000$ Pa.

\section{Nanoindentation Characterisation of the Elastic Modulus of the PDMS membrane}
\label{c:nano}

Since the Young's modulus of PDMS can significantly be altered by varying the curing temperature and time, as well as the mixing ratio of
the silicone base to the curing agent~\citep{thangawng2007, fuard08,solomon09, james10}, there is a need to measure the isotropic mechanical
properties of PDMS~\citep{mliu2009a,mliu2009b,kim2011}. Different experimental techniques have been employed to characterise
the rigidity of PDMS and the reported value of Young's modulus for PDMS usually falls within the range of 0.05--4.0 MPa~\citep{fuard08,thangawng2007}. Recently, \citet{mliu2009a} have conducted a tensile test to establish the thickness-dependent
hardness and the Young's modulus of PDMS membranes, arising due to the shear stresses that are exerted during fabrication of these thin membranes.
On the other hand, nanoindentation testing, which has been widely  employed for characterising the elastic and plastic properties of hard materials, is
also now being recognised as a tool for characterising the mechanical properties of polymeric materials. In a standard nanoindentation test, the Young's
modulus and hardness of a very thin membrane made of elastic material can easily be obtained from the load displacement
data. \citet{carrillo2005}, for example, used a nanoindentation technique to characterise the Young's modulus of PDMS with different degrees of
crosslinking.

Here, PDMS (Dow and Corning Sylgard 184) samples were prepared by mixing
the cross-linker and siloxane in a ratio of 1:10, and subsequently kept
in a vacuum chamber to remove the bubbles that were generated during mixing.
PDMS membranes with different thicknesses were then produced by spin coating
glass wafers at various rotation speeds followed by curing in an oven at 70$^{\rm o}$C for 2 hrs. The thickness of the PDMS membrane was
measured using a surface profiler. By varying the rotation speed of the spin coating process between
 $500$ and $2000$ rpm, PDMS membranes of thicknesses in the
range of 25$\mu$m to 100 $\mu$m were produced.

The nanoindentation testing is carried out using a TriboIndenter$^{\textregistered}$ (Hysitron, Inc.,
Minneapolis, MN, USA) at room temperature with a
Berkovich indenter tip. For load control function, we employ a loading and unloading rate of 10 $\mu$N/s, peak load of
100 $\mu$N and a hold period of 5 s. When the tip of the indentor reaches the sample surface, the instrument applies the predefined load and
records the load and displacement data accordingly. The hardness and the Young's modulus of the material is then determined from the
unloading portion of the load-displacement curve using classical Hertzian contact theory~\citep{johnson2003}:
\begin{equation}
H=\frac{F_{\text{max}}}{A}, \label{eq1}
\end{equation}
\begin{equation}
\frac{1}{E_r}=\frac{1-\nu^2}{E}+\frac{1-\nu_i^2}{E_i}, \label{eq2}
\end{equation}
where $H$ is the hardness of the substrate and $F_{\text{max}}$ is
the maximum force applied on the PDMS membrane. $A$ is the projected
contact area between the tip and the substrate, $\nu$ and $E$ are
the Poisson's ratio and the Young's modulus for the
test specimen, respectively, and $\nu_i$ and $E_i$ are those for the indenter. The
material properties of the diamond indenter are $E_i$ = 1140 GPa and
$\nu_i$ = 0.07. The reduced elastic modulus $E_r$ is calculated
using the following expression proposed by~\citet{oliver1992}:
\begin{equation}
E_r=\frac{S\sqrt{\pi}}{2\beta\sqrt{A}}, \label{eq3}
\end{equation}
where $S$ is the contact stiffness, taken as the initial slope of the
unloading section of the load-displacement curve, and $\beta$ is a
constant that depends on the geometry of the indenter.
\begin{figure}
\begin{center}
       \includegraphics[width=16.0cm]{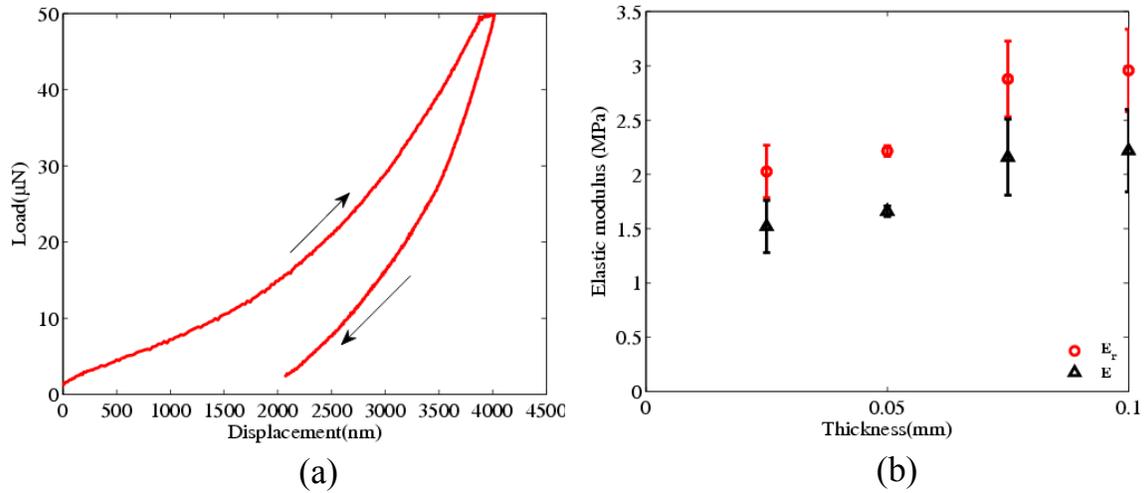}
\begin{spacing}{1.5}
        \caption{\small \label{fig1nano} (a) Load-displacement curves for a PDMS membrane with a thickness of $0.05$ mm. (b) Thickness dependence of the reduced elastic modulus $E_r$ and Young's modulus $E$ of the PDMS
membrane.}
\end{spacing}
\end{center}
\end{figure}

Figure~\ref{fig1nano}(a) shows the indentor displacement  in
response to the applied load during a nanoindentation test carried out on the PDMS
membrane using quasi-static measurements. To
confirm the reproducibility of the test data, the indentation is
performed on nine different locations of a given sample.
The penetration depth of the indenter is observed to be higher for low thickness PDMS membranes
indicating a lower value for Young's modulus as the thickness
decreases. Similar trends are also observed for other PDMS membranes.

The reduced elastic modulus $E_r$ and Young's modulus of the PDMS
membrane are calculated using Eqs.~(\ref{eq2}) and~(\ref{eq3}).
Figure~\ref{fig1nano}(b) indicates that the Young's modulus of PDMS
membranes decrease with decreasing membrane thicknesses. This is due to the polymer molecules
experiencing enhanced radial stretching due to an increase in the shear stress with the increasing
spinning speeds that are required to produce thinner membranes.
In any case, the good agreement between the
measured values for the Young's modulus therefore suggest
that the nanoindentation test is capable of differentiating the
elastic behaviour of a polymeric material with varying thickness.

\section{\label{ap:CV} Validation of the Finite-Element Formulation}

Here, we briefly provide results on the validation of the finite element formulation described in Sec. \ref{sec:ANSYSmicro} against
simple benchmark cases that have been reported earlier in the literature.

\subsection{\label{sec:CVa} Couette Flow Past a Finite Thickness Solid}

\begin{figure}
\begin{center}
       \includegraphics[width=0.50\textwidth]{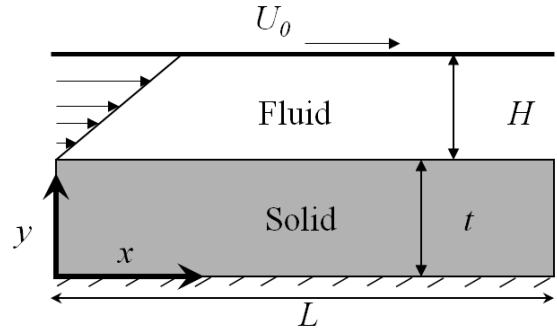}
        \caption{Schematic depiction of Couette flow of a Newtonian fluid past an incompressible finite thickness neo-Hookean solid.}
        \label{appenCfig1}
\end{center}
\end{figure}

The flow of a Newtonian fluid past an incompressible neo-Hookean solid, as shown schematically in Fig.~\ref{appenCfig1}, has been
previously described by \citet{gkanis03}. The interface between the fluid and solid is located at $y$ = $t$ and a rigid plate located at
$z$ = $(H+t)$ moves in the $x$ direction at a constant speed $U_0$, giving rise to Couette flow in the fluid domain; the
bottom edge of the solid is held fixed. \citet{gkanis03} performed a linear stability analysis of this problem in the limit of zero Reynolds
number Re and infinite domain length $L$, and have shown that the steady-state solution of the deformation in the solid produced by
the Couette flow is
\begin{equation}
\boldsymbol{x}=(x, y) = (X+\Gamma Y, Y),
\end{equation}
where $\Gamma$ is the dimensionless number defined as $\Gamma = (\eta U_{0})/(G H)$.

\begin{figure}
\begin{center}
       \includegraphics[width=0.90\textwidth]{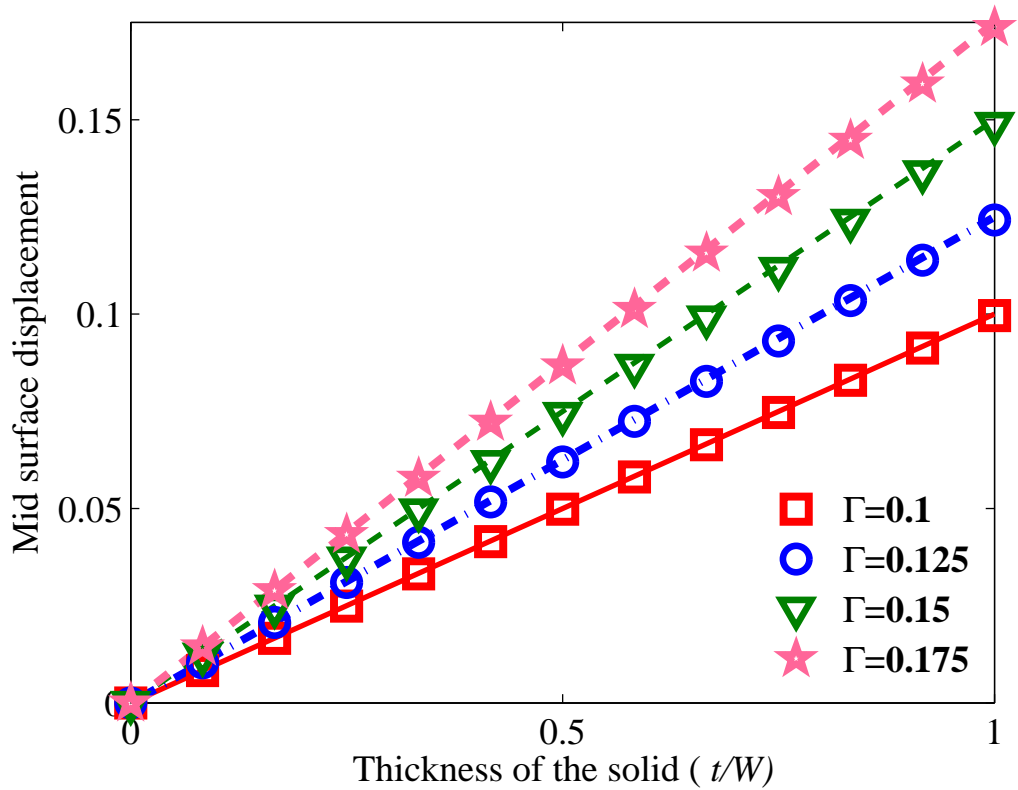}
        \caption{Comparison of the finite element simulation with the analytical solution for different values of $\Gamma$. The lines denote the analytical
        solution reported by~\citet{gkanis03} whereas symbols are the predictions from  the our simulation.}
        \label{appenCfig2}
\end{center}
\end{figure}

Computations have been performed to compare predictions for the deformation of the solid domain at $L/2$ with the analytical results
of~\citet{gkanis03}. In order to eliminate end effects caused by the fixed ends of the solid and fluid domains in computations, we have varied the
length of the domain between 10 and 30~m and ensured that domain length independent predictions are obtained.
The following parameter values have been used: $\rho = 10^{-3}$ kg/m$^3$, $\eta = 1\ {\rm Pa}\cdot{\rm s}$, $H = 1$ m, $t = 1$ m and
$U_0 = 1 \times 10^{-3}$--$1.75 \times 10^{-3}$ (such that Re $\sim 0$) and $G = 10^{-2}$ Pa. This choice of parameter values maintains
$\Gamma$ in the range $0.1$--$0.175$.
The mid-surface displacement of the solid predicted by the finite-element formulation is compared with the analytical solution
for different values of $\Gamma$. Figure~\ref{appenCfig2} shows that in all
cases the predictions of our finite element simulation are in excellent agreement with the analytical solution.

\subsection{\label{sec:CVb} Flow in Two-Dimensional Collapsible Channels: Elastic Beam Model}

\citet{luo07} have carried out extensive studies of Newtonian fluid flow
in a two-dimensional collapsible channel by considering the flexible wall to be
a plane-strained elastic beam that obeys Hooke's law. In contrast to the
current finite thickness elastic solid model, the beam model does not admit any stress
variation across the beam cross-section.

For the purposes of comparison, the dimensions of the channel and other parameter values are chosen to be identical to those used
by~\citet{luo07} in their simulations: $L_u = 5H$, $L = 5H$, $L_d = 30H$,
$U_0 = 0.03$ m/s, $H = 10^{-2}$ m, $\rho = 10^3$ kg/m$^3$ and $\eta = 0.001\ {\rm Pa}\cdot{\rm s}$. This choice corresponds to Re $=300$. Further, we set $G = 11.97$ kPa (which is equivalent to a value of
$35.9$ kPa for the Young's modulus of an incompressible solid) and $p_e = 1.755$ Pa. The flexible wall thickness is varied in the range $0.01H$--$0.1H$. We note that the `pre-tension' in the beam is also a variable in the model of
\citet{luo07}; however, since no such variable exists in the current model, we have restricted
our comparison to the results reported by~\citet{luo07} for cases where the bream pre-tension is zero.

\begin{figure}
\begin{center}
       \includegraphics[width=0.90\textwidth]{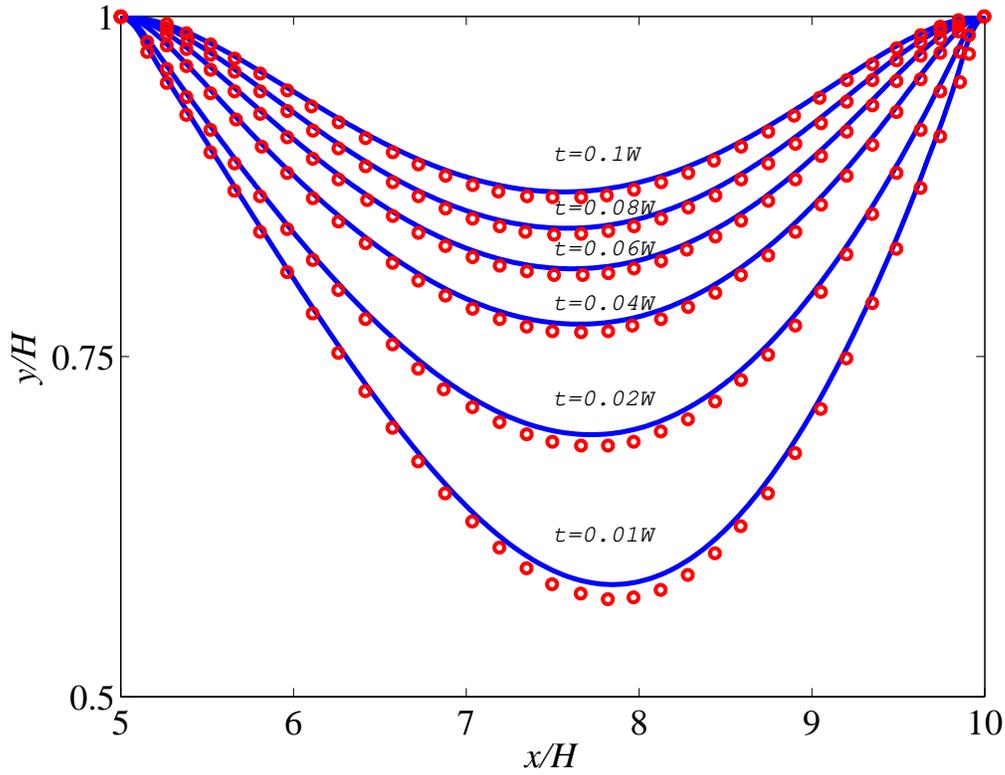}
        \caption{Comparison of the shape of the flexible wall predicted by
        the finite thickness elastic solid model
        with the results of~\citet{luo07} for an elastic beam model with different values of the wall thickness.
        Circles denote the elastic beam model whereas lines denote the results of our numerical simulation.}
        \label{appenCfig3}
\end{center}
\end{figure}

Figure~\ref{appenCfig3} compares the prediction of the shape of the flexible wall of our finite thickness elastic solid model with that reported by
of~\citet{luo07}. While our simulations agree with~\citet{luo07} for the relatively small
deformations that occur at large membrane thicknesses $t$, the Hookean beam model, as expected, begins to depart from the prediction of the
nonlinear neo-Hookean model for  large deformations that are associated with small
membrane thicknesses.

\subsection{\label{sec:CVzm}  Flow in Two-Dimensional Collapsible Channels: Zero-Thickness Membrane Model}

Simulations have also been performed to compare predictions of the flexible wall shape by
current finite thickness elastic solid model with that of a zero-thickness
membrane model of~\citet{luo95} for the flow of a Newtonian fluid. Apart from the
simplicity of the zero-thickness membrane model from a constitutive point of view, a fundamental
difference between the two models is that while the tension in the
flexible wall is prescribed \textit{a priori} in the zero-thickness membrane model, it is part of the solution
in the finite thickness elastic solid model. As a result, a several step procedure is required to carry out the comparison, as described in
what is to follow.
\begin{figure}
\begin{center}
       \includegraphics[width=8.0cm]{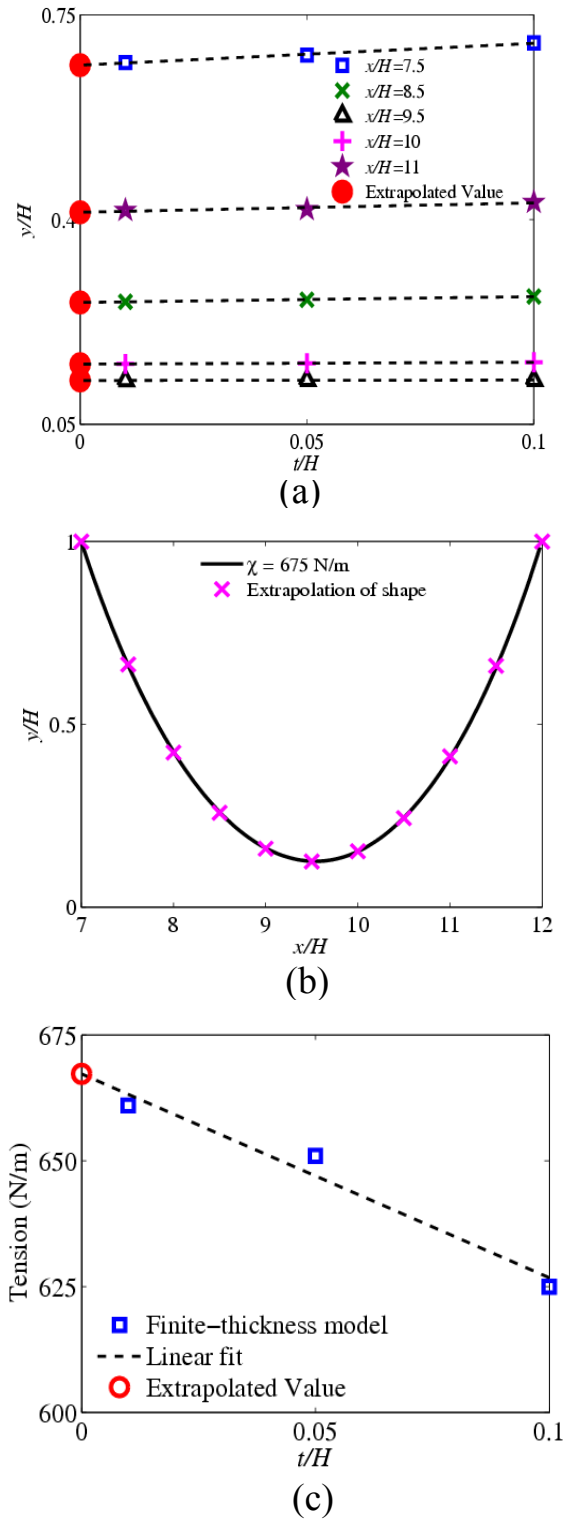}
\begin{spacing}{1.5}
        \caption{\small  \label{appenCfig5} (a) Extrapolation to $t = 0$ of the
flexible wall shape  obtained from the finite thickness elastic solid
model for $t = 0.01H$, $t = 0.05H$, {and} $t = 0.1H$.
(b) Comparison of the shape of the flexible wall predicted by the
finite-thickness solid model (symbols) with the prediction of the zero-thickness
membrane model (solid line). (c) Extrapolation of the average tension acting
in the elastic solid to the limit of zero wall thickness.}
\end{spacing}
\end{center}
\end{figure}

The zero-thickness membrane model is first computed for a pre-determined value of
membrane tension equal to $675$ N/m, with the following parameter values:
Re $ = 1$, $\rho = 1054$ kg/m$^3$, $U_0 = 1.338 \times 10^{-2}$ m/s, $H = 10^{-2}$ m, $\eta = 0.141\ {\rm Pa}\cdot{\rm s}$ and $p_e = 17545$
N/m$^2$. This leads to a prediction of the minimum height of the gap in the channel (beneath
the flexible membrane) of  $h/H = 0.125$. Computations with the finite thickness elastic solid model are then carried out for the same parameter values,
for various combinations of flexible wall thickness $t$  and shear modulus $G$ such that each combination always
leads to the same value of the minimum channel gap height, namely $h/H = 0.125$. It turns out that even though the
minimum gap height is the same in both models, the predicted interface shape is not, with the difference increasing as
the thickness of the elastic solid increases. This is clearly a result of the finite thickness of the elastic solid. Consequently, in order to compare the interface shape, we carry out an extrapolation procedure in which the interfacial height of the interface at various locations in the gap as a function of the flexible wall thickness is extrapolated to the limit of zero wall thickness, as shown in Fig.~\ref{appenCfig5}(a). The extrapolated interface shape is then compared with the prediction by the zero-thickness membrane model in Fig.~\ref{appenCfig5}(b), in which we observe excellent agreement between the two models.

The remaining step involves the evaluation of the resultant tension in the finite thickness elastic solid and how it compares with the pre-determined
membrane tension of $675$ N/m. Here, we first estimate the tension in the finite thickness solid at a particular location $x$ by averaging the tangential
solid stresses acting across the cross-section at $x$. An estimate of the overall tension in the solid is then obtained by averaging the tension
along the entire length of the flexible solid for all values of $x$. The values of the average tension obtained from the finite thickness
elastic solid model for $t = 0.01H$, $t = 0.05H$ and $t = 0.1H$ are then extrapolated to $t = 0$, as shown in Fig.~\ref{appenCfig5}(c), in which we
observe the extrapolated value of tension to be fairly close to the value of $675$ N/m used in the zero-thickness membrane model.

\end{document}